\documentclass{article}

\usepackage{graphicx}
\usepackage{amsmath}
\usepackage{times}
\usepackage{natbib}
\usepackage{xspace}
\usepackage{emulateapj}

\begin{document}

\title{Quasi-Periodic Oscillation in Seyfert galaxies: Significance
  levels. \\ The Case of Mrk~766}

\author{
Sara Benlloch\altaffilmark{1},
J\"orn~Wilms\altaffilmark{1},
Rick Edelson\altaffilmark{2},
Tahir Yaqoob\altaffilmark{3,4},
R\"udiger Staubert\altaffilmark{1},
}

\altaffiltext{1}{Institut f\"ur Astronomie und
Astrophysik--Astronomie, University of T\"ubingen, Waldh\"auser
Stra\ss{}e 64, D-72076 T\"ubingen, Germany}

\altaffiltext{2}{University of California, Los Angeles, Department of Astronomy, Los Angeles, CA 90095-1562}

\altaffiltext{3}{Laboratory for High Energy Astrophysics, NASA Goddard
  Space Flight Center, Greenbelt, MD 20771}

\altaffiltext{4}{Johns Hopkins University, Department of Physics and
  Astronomy, Homewood Campus, 3400 North Charles Street, Baltimore, MD
  21218} 

%\email{benlloch@astro.uni-tuebingen.de}

\begin{abstract} 
  We discuss methods to compute significance levels for the existence of
  quasi-periodic oscillations (QPOs) in Active Galactic Nuclei (AGN) which
  take the red-noise character of the X-ray lightcurves of these objects
  into account. Applying epoch folding and periodogram analysis to the
  \textsl{XMM-Newton} observation of the Seyfert galaxy Mrk~766, a possible
  QPO at a timescale of 4200\,s has been reported.  Our computation of the
  significance of this QPO, however, shows that the 4200\,s peak is not
  significant at the 95\% level.  We conclude that the 4200\,s feature is
  an artifact of the red-noise process and not the result of a physical
  process within the Active Galactic Nuclei.
\end{abstract}


%\clearpage

\section{Introduction}

X-ray quasi-periodic oscillations (QPOs) are among the most important
observational properties of galactic X-ray binaries \citep[XRBs; see][for a
recent review]{vanderklis:00a}, yielding important constraints on the mass
of the central black hole (BH), $M_{\text{BH}}$, and providing theoretical
clues and constraints on the operative physical processes and geometry in
the regime of strong gravity.  Although their origin is still a matter of
scientific debate \citep[][and references therein]{psaltis:00a}, there is
general agreement that QPOs originate close to the central BH.
 
To date, QPOs at timescales of a few kiloseconds have also been claimed for
some active galactic nuclei (AGN): NGC~4151 \citep{fiore:89a}, NGC~6814
\citep{mittaz:89a}, NGC~5548 \citep{papadakis:93b}, NGC~4051
\citep{papadakis:95b}, RX~J0437.4$-$4711 \citep{halpern:96a},
IRAS~18325$-$5926 \citep{iwasawa:98a}, MCG$-$6-30-15 \citep{lee:00a},
Mrk~766 \citep[][]{boller:01a}, and IRAS~13224$-$3809
\citep{pfefferkorn:01a}. In addition, possible long term periodicities with
periods of months have been claimed for the radio loud AGN Mrk~421,
Mrk~501, and PKS~2155$-$304 \citep{osone:01a}.  Some of the kilosecond QPO
findings, however, are controversial, with the strongest EXOSAT result
(NGC~5548) being disputed by \citet{tagliaferri:96a}, who attribute it
mostly to periodic swapping of detectors.  NGC~6814 turned out to be
confused with a cataclysmic variable \citep{madejski:93a,staubert:94a}.

Given the important implications based upon detection of QPO in AGN, it
seems worthwhile to study the methods employed to determine their
significance in detail. In this \textit{Letter}, we present a study of how
to compute this significance using Monte Carlo simulations of lightcurves
with the method of \citet{timmer:95a} and using periodogram analysis and
epoch folding to detect the periodicity (\S\ref{sec:qpo}).  We then apply
our methods to a reanalysis of the \textit{XMM-Newton} lightcurve of
Mrk~766 and find that the QPO claimed in \citet{boller:01a} has in fact low
statistical significance (\S\ref{sec:mkn}).  In \S\ref{sec:disc} we discuss
our results and comment on further work.

\begin{figure*}
  \centerline{\includegraphics[width=0.95\textwidth]{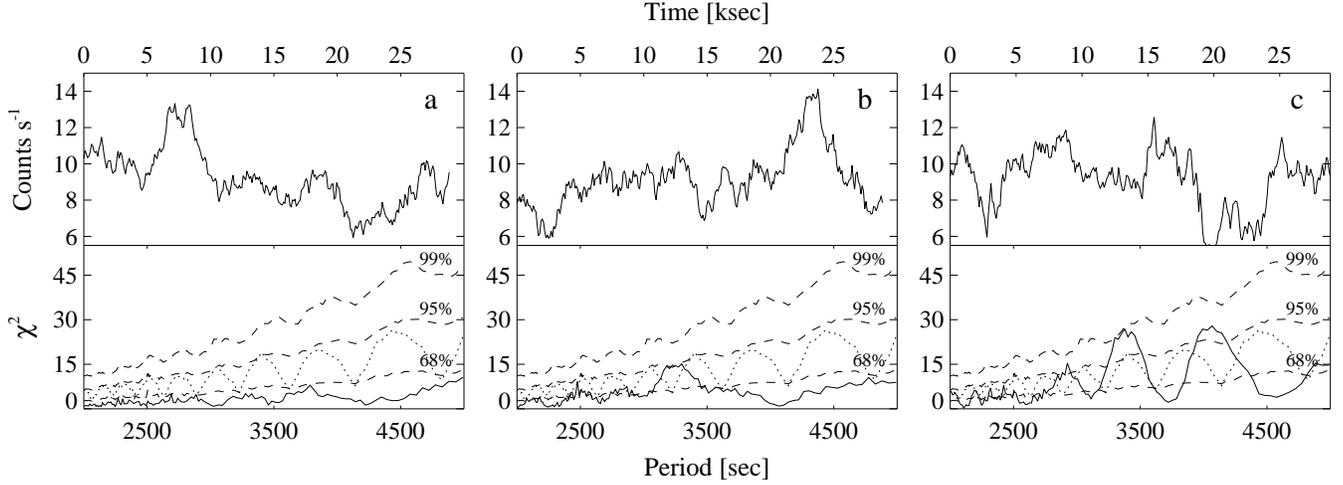}}
  \figcaption{Three examples of typical red-noise lightcurves for a process
    with an {\scriptsize{$f^{-1.9}$}} spectrum binned to a resolution of
    100\,s using the method described by \citet{timmer:95a}.  \textit{Upper
      panels}: Simulated red-noise lightcurves.  \textit{Lower panels}:
    {\scriptsize{$\chi^2(P)$}} curves from epoch folding (solid line) of
    the corresponding simulations. The dashed lines represent the 99\%,
    95\%, and 68\% ``local significance'' levels (see text for definition)
    obtained for a sample of 5000 Monte Carlo simulations using the
    \citet{timmer:95a} method. The dotted line represents the 99\%
    significance level obtained with the ``phase randomization'' method,
    which overestimates the significance of the peaks in the
    {\scriptsize{$\chi^2(P)$}} curves. Note that the presence of one or
    more peaks in the {\scriptsize{$\chi^2(P)$}} curve is far from unusual
    for red-noise lightcurves.\label{fig:sim_plots}}
\end{figure*}

\section{Significance of QPO Detections}\label{sec:qpo}

Currently, two methods for the variability analysis of astronomical
sources, especially in searching for periodic signals, are common in
astronomy: periodogram analysis and epoch folding. We only give a brief
description of these methods here, see, e.g., \citet{vanderklis:89a},
\citet{leahy:83a} and \citet{davies:90a} for in-depth discussions.  Based
on the Fourier decomposition of the lightcurve, periodogram analysis (often
called power spectrum density analysis, PSD) is especially sensitive to
periodic signals with a modulation that is close to sinusoidal.  On the
other hand, epoch folding ($\chi^2(P)$) is based on comparing pulse
profiles obtained from binning the data into phase bins at a test period,
$P$, with a constant count rate using a $\chi^2$ test.  In both methods,
the detection of a periodic signal is claimed if the PSD value or the
$\chi^2$ value at the period of interest is significantly above the values
of the testing statistics surrounding this period.

In order to safely use either of these methods, it is critically important
to accurately estimate the significance of QPO features. This is not
trivial for real AGN lightcurves, which have much worse sampling and signal
to noise than X-ray binary lightcurves.  Were the X-ray lightcurve purely
dominated by white noise and evenly sampled, the significance could be
easily determined from the statistical properties of the PSD or from the
$\chi^2(P)$ statistics.  This assumption, however, does not apply to AGN,
where the PSD can be well approximated by a power-law $P(f)\propto
f^{-\beta}$ with $\beta\sim 1\ldots 2$ for the frequencies of interest
\citep{lawrence:93a,green:93a,koenig:97a,edelson:99a}.

When studying a feature in $\chi^2(P)$ or a PSD, we can ask two different
questions: 1.~What is the significance of a QPO peak at a given
(predefined) frequency (the ``local significance'' of the QPO), and,
2.~what is the significance of a QPO feature seen in a given frequency
range (the ``global significance'' of the QPO). Note that these questions
are really different questions, since we do have more knowledge about the
QPO in the case of the ``local significance'' (we do know its frequency),
while in the case of the ``global significance'', we are only interested in
having the question answered that a feature somewhere in the interesting
frequency range is significant or not. To our knowledge, however, no
formulae for determining the significance of a peak in the PSD or in
$\chi^2(P)$ exist for such red-noise lightcurves.  A common approach,
therefore, is to resort to Monte Carlo simulations \citep[see,
e.g.,][]{horne:86a}.  In these simulations, a large number ($\gtrsim 1000$)
of lightcurves with the same statistical properties and same temporal
sampling as the original lightcurve are generated and their $\chi^2(P)$ is
computed.

For computing the ``local significance'', one determines the statistical
distribution of the resulting $\chi^2(P)$ value at the frequency of
interest. A deviation of the measured $\chi^2(P)$ from the red-noise
$\chi^2(P)$ is significant if it is above a certain threshold determined
from this distribution. Typically, threshold values of 99\% or even 99.9\%
are used \citep{bevington}. An identical procedure can be used to test the
significance of a peak in the PSD.

For determining the ``global significance'', a similar approach is used.
Since we are looking at $\chi^2(P)$ values determined at different periods,
one has to take the red-noise character of the data and window effects into
account. Instead of directly using $\chi^2(P)$ values, more robust tests
can be devised by first dividing the computed individual $\chi^2(P)$ values
by the average $\langle\chi^2(P)\rangle$, obtained by averaging the
$\chi^2(P)$ curves of the simulated lightcurves. To determine the ``global
significance'' of the maximum $\chi^2(P)$-peak of an observed lightcurve,
we therefore propose to compare this value to the distribution of $\xi_{\rm
  max}:=\max\left\{ \chi^2(P)/\langle \chi^2(P)\rangle\right\}$ from the
simulated lightcurves. Here, the maximum is to be taken over the period
range of interest.  We would consider an observed peak as likely due to a
physical effect only if it is above the 99\% threshold determined from the
distribution. The maximum $\chi^2(P)$-peaks present in the red-noise
lightcurves of Fig.~\ref{fig:sim_plots}b and~c have a ``global
significance'' of 61\% and 80\% respectively.  Similar global tests can
also be devised for the distribution of the 2nd or 3rd largest $\chi^2(P)$
value and for the significance of the largest peaks in a PSD, although we
will not use them here.

It is crucially important how lightcurves are produced by Monte Carlo
simulations, if the significance level of any peak seen in a red-noise PSD
is to be determined. One usually starts with a model PSD and performs an
inverse Fourier transformation to obtain a lightcurve.  An often used
algorithm, called the ``phase randomization'' method \citep{done:92a},
determines the Fourier amplitude from the square root of the power-law
shaped PSD and assumes the Fourier phase to be uniformly distributed in
$[0,2\pi[$. Although the PSDs produced by ``phase randomization'' are
$\propto f^{-\beta}$, it was pointed out by \citet{timmer:95a} and
\citet{papadakis:95b} that the resulting lightcurves do not resemble the
pure red-noise process in all of their statistical properties.  First, this
procedure chooses a deterministic amplitude for each frequency and only
randomizes the phases.  All simulated lightcurves thus exhibit a trend
caused by the dominating lowest frequency.  Secondly, the periodogram of a
red-noise lightcurve must obey the usual periodogram statistics: the PSD
follows approximately a $\chi^2$ distribution with two degrees of freedom,
$\chi^2_2$, i.e., the standard deviation of each PSD point is of the same
magnitude as the PSD value itself such that the periodogram is fluctuating
wildly \citep[see, e.g.,][]{vanderklis:89a}.  ``Phase randomization'' does
not take into account this randomness of the periodogram according to the
$\chi^2_2$ distribution and therefore the uncertainty of the related
distribution of the estimated periods is significantly underestimated. In
other words, red-noise PSDs have a larger scatter than those obtained from
``phase randomization'' -- including the possibility of outliers that
strongly deviate from the general $f^{-\beta}$ behavior.  This is important
since these outliers could be interpreted as quasi periodic oscillations,
while in reality they result from the statistics of the red-noise process.
We note that the frequent occurrence of random peaks in red-noise PSDs is
avoided in X-ray binary work by averaging many PSDs to obtain the ``true
PSD'' of a source \citep{nowak:99b,vanderklis:89a}. For AGN, averaging the
PSDs is unfortunately not possible due to the prohibitively long
observation times required.

As we have mentioned above, the ``phase randomization'' algorithm will
not produce lightcurves with the appropriate red-noise statistical
characteristics.  Indeed, were one to use the Monte Carlo approach outlined
above to lightcurves computed with ``phase randomization'', the QPO
significance would be overestimated. Instead, an algorithm that produces
lightcurves with the correct statistical red-noise behavior has to be used.
For this purpose \citet{timmer:95a} proposed a new algorithm.  This
algorithm \citep[used also by][]{green:99a} allows randomness both in phase
and in amplitude producing the desired $\chi^2_2$ distribution in the
periodogram. In the following, we will use the \citet{timmer:95a} algorithm
to simulate red-noise lightcurves.  As an illustration, three simulated
examples of red-noise lightcurves with significance levels obtained from
the Monte Carlo simulations are displayed in Fig.~\ref{fig:sim_plots}. We
also display the significance levels obtained with ``phase randomization''
to indicate that these levels are in fact much smaller than the correct
levels and therefore would imply an overestimate in the significance of an
apparent period.

We note that Poisson noise introduces additional ``observational noise'' in
the measured lightcurves. This observational noise has to be added to the
simulated lightcurves after the inverse Fourier transform has been
performed. For each time bin, the number of observed photons is drawn from
a Poisson distribution with its mean given by $r(t) \Delta t$, where $r(t)$
is the simulated count rate and $\Delta t$ is the binning.  For data with
new instruments such as \textit{XMM-Newton}, with a high signal to noise
ratio, this latter step can be ignored if the source is bright enough. For
example, with $\Delta t=100$\,s, ``observational noise'' contributes less
than 3\% to the Mrk~766 lightcurve. For earlier instruments such a
simplification of the Monte Carlo algorithm is not possible.  We note that
in earlier work employing ``phase randomization'' and the addition of
observational noise, the resulting PSD statistics would asymptotically
approach the $\chi^2_2$ distribution and therefore the overestimation of
the period significance would be less than with newer data with a high
signal to noise ratio.

\centerline{\includegraphics[width=0.45\textwidth]{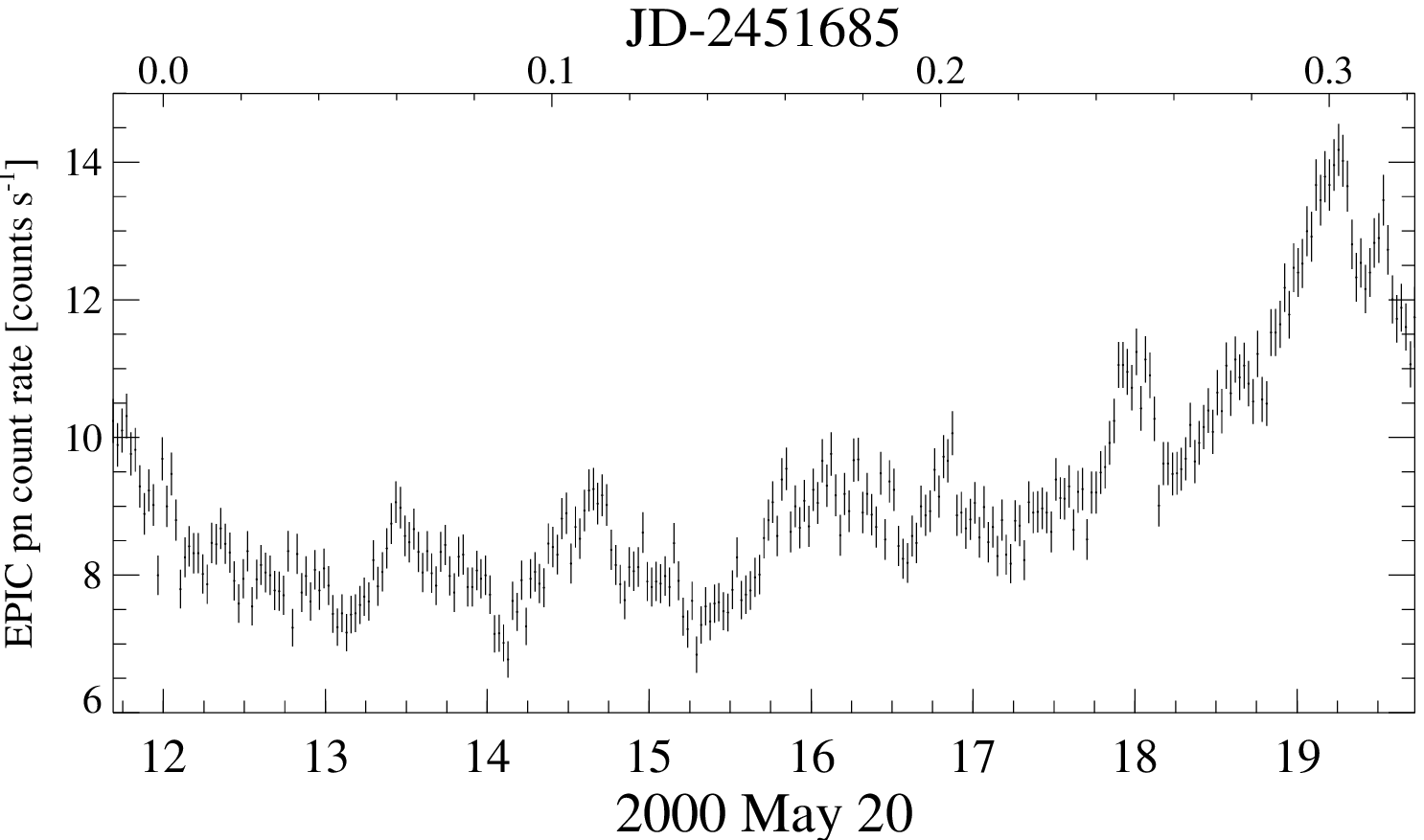}}
\figcaption{XMM-Newton EPIC-pn lightcurve of Mrk~766 for the 0.2--2\,keV
  band with a resolution of 100\,s.\label{fig:lc}}

\section{The Case of Mrk~766}\label{sec:mkn}

We now apply the methods outlined in the previous section to the putative
QPO in \textit{XMM-Newton} data of Mrk~766 taken during revolution 0082. We
concentrate on data from the EPIC-pn instrument \citep{strueder:01a} which
was operated in the small window mode during the observation.

\subsection{Data Extraction}

We extracted source photons from a circle of 9.5 pixels radius centered on
the source (detector coordinates (37.5,54)). For the background, data from
an off-axis position (18,17.5) extracted with a circle of then same radius
were used.  The time range of the lightcurve was chosen to be consistent
with the approach of \citet{boller:01a} and results in an exposure time of
29\,ksec. After background subtraction, we corrected the measured
count-rates as is appropriate for the $\sim$71\% live-time during the
5.7\,ms readout cycle of the pn-CCD \citep{kuster:99a}.  Fig.~\ref{fig:lc}
displays the resulting lightcurve.

\subsection{Lightcurve Analysis}

We display the PSD and the $\chi^2(P)$ curves in Fig.~\ref{fig:sig} (the
frequency range is $0.4 \times 10^{-4}$\,Hz to $30 \times 10^{-4}$\,Hz with
57 independent frequencies, the period range is 2000--5000\,s with 132 test
periods).  A peak at a period of $\sim$4200\,s (corresponding to a
frequency of $\sim 2.5 \times 10^{-4}$\,Hz) that is consistent with the
period claimed by \citet{boller:01a} is seen. In order to determine the
significance of the peaks seen in Fig.~\ref{fig:sig}, we computed
significance levels using the methods outlined in section~\ref{sec:qpo}.

Before we can perform these simulations, we need to determine the shape of
the PSD. For this purpose, we apply the ``response method'' of
\citet{done:92a} and \citet{green:99a}, where a model power spectrum
generated through the combination of a large number of red noise simulated
lightcurves is compared to the observed power spectrum.  Varying the slope
and normalization of the model periodogram, the best $\chi^2$ fit gives a
slope of $\beta=1.9$, and a normalization determined by the variance of the
original XMM lightcurve ($\chi^2$/dof = 17.89/55).  This low value
indicates that the observed PSD is fully consistent with a power law and
that no additional components, such as a QPO, are required.

\vspace*{0.2cm}
\centerline{\includegraphics[width=0.45\textwidth]{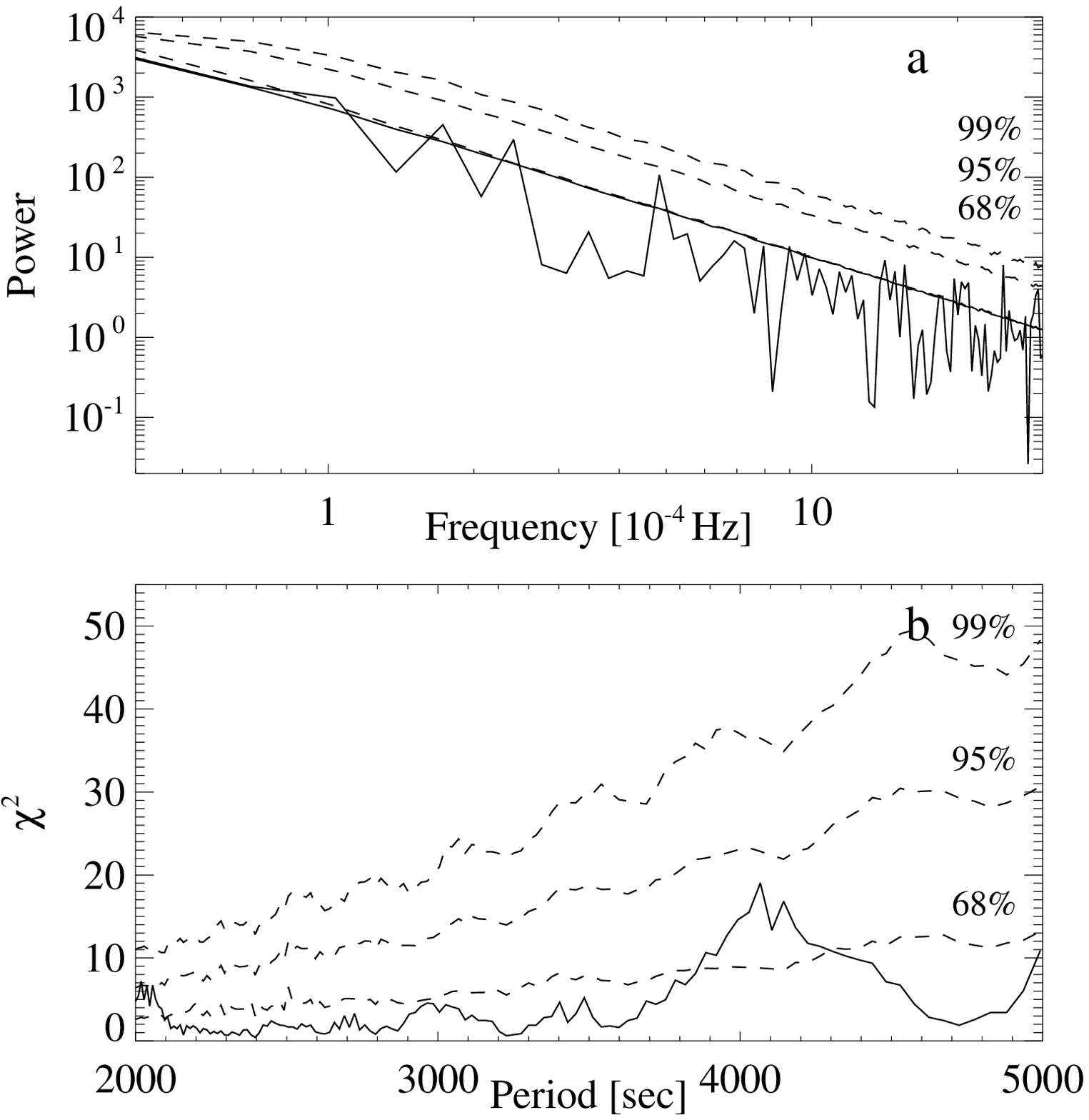}}
\figcaption{\textit{a} PSD in \citet{leahy:83a} normalization and best
  power-law fit using the \citet{done:92a} method (continuous lines) and,
  \textit{b} {\scriptsize{$\chi^2(P)$}} curve (continuous line) of the XMM
  Mrk~766 lightcurve of Fig.~\ref{fig:lc}.  The set of dashed lines in both
  panels represent in ascending order the $68\%$, $95\%$, and for $99\%$
  ``local significance'' levels for a set of 5000 Monte Carlo red-noise
  simulations with $\beta =1.9$.\label{fig:sig}} \vspace*{0.2cm}

We note, however, that the short duration of the observation only allows
for a poor determination of the PSD shape. Methods complementary to testing
the consistency of the PSD with a power law should be used.  We therefore
simulated the desired 5000 lightcurves from a $f^{-1.9}$-PSD, with 29000\,s
of duration using a sampling interval of 100\,s, a mean count rate
$\mu=9.15\,\text{counts}\,\text{s}^{-1}$, a lightcurve variance
$\sigma^2=2.46\,\text{counts}^2\,\text{s}^{-2}$.  Examples of such
simulated red-noise lightcurves are shown in Fig.~\ref{fig:sim_plots}.

To obtain the ``local significance'' of the peak at $\sim$4200\,s , we
calculated $\chi^2(P)$ and the PSD of each of the 5000 Monte Carlo
realizations.  For each trial period, we then computed the distribution of
the $\chi^2(P)$ and PSD values at this trial period from all realizations.
These distributions were used to determine the 99\%, 95\%, and 68\% ``local
significance'' levels using the method described in section~\ref{sec:qpo}.
Our results are shown in Fig.~\ref{fig:sig}.  The peak at $\sim$4200\,s
claimed as QPO by \citet{boller:01a} is below the 95\% significance curve.

\vspace*{0.2cm}
\centerline{\includegraphics[width=0.45\textwidth]{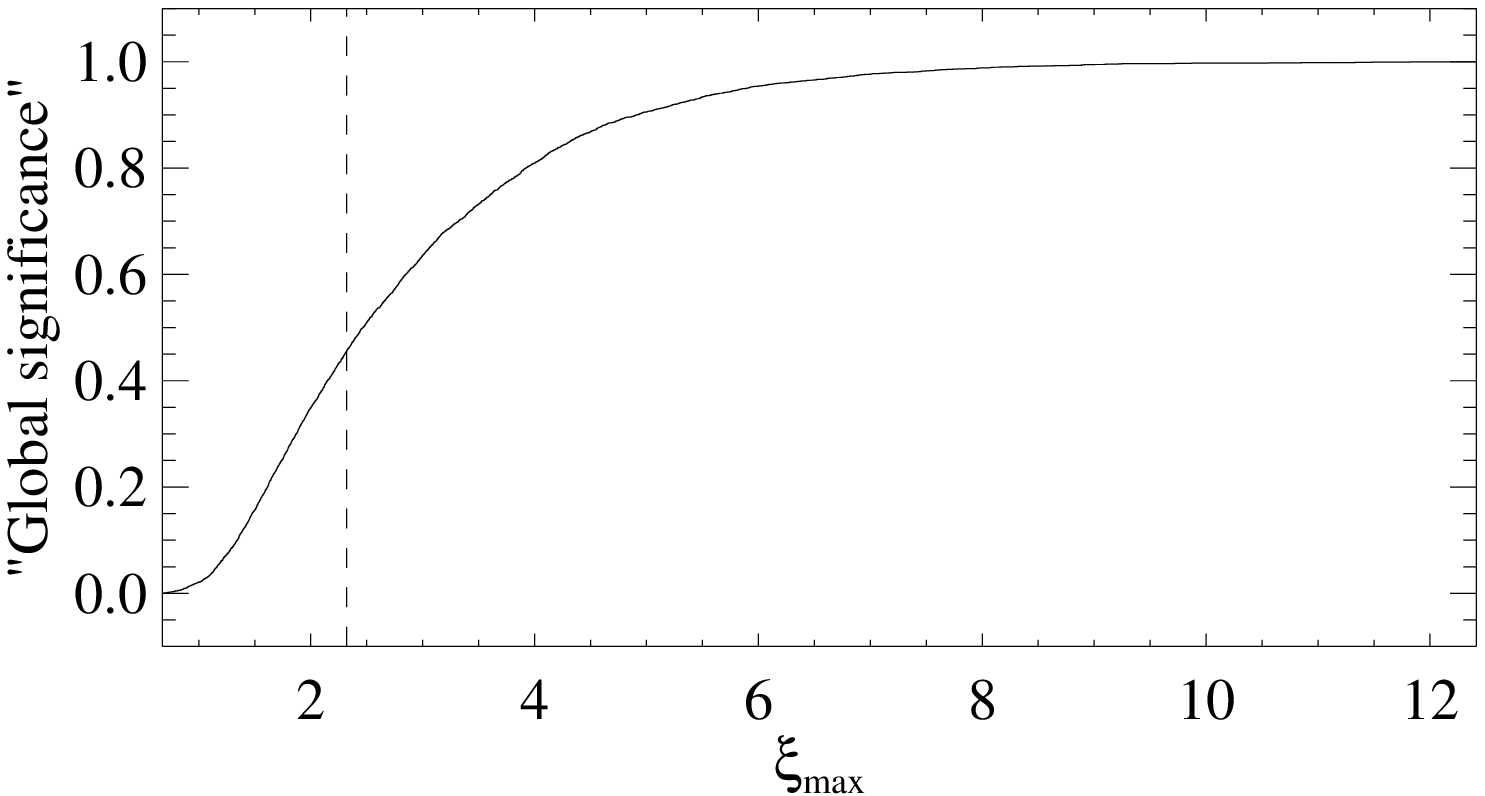}}
\figcaption{Probability distribution of the {\scriptsize{$\xi_{\rm max}$}}
  values (see text for definition) from the 5000 red-noise simulated
  lightcurves in the 2000--5000\,s period range. The ``global
  significance'' indicates the probability of finding a lightcurve with a
  maximum peak with a value greater than the correspondent
  {\scriptsize{$\xi_{\rm max}$}} value.  The dashed line marks the position
  of the {\scriptsize{$\xi_{\rm max}$}} value of the 4200\,s feature in
  Mrk~766 in the 2000--5000\,s period range.  The ``global significance''
  of this peak corresponds to a value of 45\%.
  \label{fig:rel_sig}}
\vspace*{0.2cm}

We also applied the ``global significance'' test to the Mrk~766 lightcurve
for the period range from 2000\,s to 5000\,s.  Fig.~\ref{fig:rel_sig} shows
the probability distribution of $\xi_{\rm max}$, which corresponds to the
probability of finding a red-noise lightcurve with a normalized maximum
peak value less than or equal to the corresponding $\xi_{\rm max}$ value.
For Mrk~766, the putative QPO peak lies at the 45\% mark. In other words,
$\sim$50\% of the simulated red-noise lightcurves show peaks that are more
significant than the peak observed in Mrk~766. We conclude that the
observed 4200\,s ``QPO'' in Mrk~766 may be an artifact of the red-noise
process and not the result of a physical process.

As we mentioned above, the value of $\beta$ determined from the
observations is rather uncertain. However, our result is independent of the
specific value of $\beta$: Simulations with $\beta$ values ranging from 1
to~2, i.e., over the typical $\beta$ range seen in AGN, yielded similar
results.

\section{Conclusions}\label{sec:disc}
In this \textit{Letter} we have described two methods to determine
the significance of possible quasi-periodic signals in the red-noise
lightcurves observed from Active Galactic Nuclei using Monte Carlo
simulations; a frequency-dependent ``local significance'' test and a
``global significance'' test.  Reiterating arguments by \citet{timmer:95a},
we showed that ``phase randomization'' techniques should not be used for
the generation of simulations since the resulting lightcurves do not
exhibit true red-noise characteristics. The periodograms of lightcurves
produced by this method do not obey the $\chi^2$ statistics which results
from a random process, overestimating therefore the significance of peaks
present in the periodogram of an underlying red-noise lightcurve.  Instead
of ``phase randomization'' we recommend the algorithm of \citet{timmer:95a}
to simulate red-noise lightcurves with the correct statistical properties.

The ``local significance'' test based on red-noise power simulations
generated with the \citet{timmer:95a} algorithm shows that the 4200\,s
feature in Mrk~766 is not significant at the 95\% level.  This statement
holds for both, PSD and $\chi^2(P)$ analysis (Fig.~\ref{fig:sig}). The
``global significance'' of the 4200\,s feature is 45\%, i.e., higher peaks
in the 2000--5000\,s period range are found in roughly half of all simulated
lightcurves. Thus, the presence of a peak in a red-noise $\chi^2(P)$ or PSD
is far from unusual.  We are therefore led to the conclusion that we cannot
confirm the claim of a $\sim$4200\,s QPO.  Rather, we attribute the peak at
4200\,s to a random occurrence which is due to the red-noise character of
AGN data.

We are currently in the process of checking the kilosecond QPO detections
claimed for AGN using archival data and the methods outlined in this
\textit{Letter}. We will report on our results in a forthcoming
publication.

\acknowledgments

We thank R.E.~Rothschild for useful discussions.  We acknowledge support
from DLR grant 50~OX~0002, and NASA grants NAG~5-7317, NAG~5-9023, and
NCC5-447. This work is based on observations obtained with
\textit{XMM-Newton}, an ESA science mission with instruments and
contributions directly funded by ESA Member States and the USA (NASA).

%\clearpage

\end{document}